\documentclass[twocolumn,preprintnumbers,superscriptaddress,amsmath,amssymb,nofootinbib]{revtex4}

\usepackage{graphicx}
\usepackage{dcolumn}
\usepackage{bm}
\usepackage{amsmath, amssymb}

\usepackage[normalem]{ulem}  

\renewcommand\sout{\bgroup \color{red} \ULdepth=-.5ex \ULset}

\newcommand{\Psfig}[2]{\includegraphics[width=#1]{#2}}
\newcommand{\PsfigII}[2]{\includegraphics[scale=#1]{#2}}

\def\Eglab{E_{\gamma}^{\rm lab} }
\def\deutwf{\tilde{\varphi} }
\def\eprime{\eta ^{\prime} }
\def\eboth{\eta ^{( \prime )} }

\def\mev{\text{ MeV}}
\def\gev{\text{ GeV}}

\def\nanobsr{\text{ nb}/\text{sr}}

\allowdisplaybreaks

\begin{document}

\preprint{}

\title{\boldmath Possible $\eta ^{\prime} d$ bound state and its
  $s$-channel formation in the $\gamma d \to \eta d$ reaction}

\author{Takayasu Sekihara} 
\email{sekihara@post.j-parc.jp}
\affiliation{Advanced Science Research Center, Japan Atomic Energy
  Agency, Shirakata, Tokai, Ibaraki, 319-1195, Japan}

\author{Hiroyuki Fujioka} 
\altaffiliation[Present address: ]{Department of Physics,
  Tokyo Institute of Tech- nology, Tokyo 152-8551, Japan.}
\affiliation{Department of Physics, Kyoto University, Kyoto 606-8502,
  Japan}

\author{Takatsugu Ishikawa}
\affiliation{Research Center for Electron Photon Science (ELPH),
  Tohoku University, Sendai 982-0826, Japan}

\date{\today}

\begin{abstract}

  We theoretically investigate a possibility of an $\eprime d$ bound
  state and its formation in the $\gamma d \to \eta d$ reaction.
  First, in the fixed center approximation to the Faddeev equations we
  obtain an $\eprime d$ bound state with a binding energy of $25 \mev$
  and width of $19 \mev$, where we take the $\eprime N$ interaction
  with a coupling to the $\eta N$ channel from the linear $\sigma$
  model.  Then, in order to investigate the feasibility from an
  experimental point of view, we calculate the cross section of the
  $\gamma d \to \eta d$ reaction at the photon energy in the
  laboratory frame around $1.2 \gev$.  As a result, we find a clear
  peak structure with the strength $\sim 0.2 \nanobsr$, corresponding
  to a signal of the $\eprime d$ bound state in case of backward
  $\eta$ emission.  This structure will be prominent because a
  background contribution coming from single-step $\eta$ emission off
  a bound nucleon is highly suppressed.  In addition, the signal can
  be seen even in case of forward $\eta$ emission as a bump or dip,
  depending on the relative phase between the bound-state formation
  and the single-step background.
  
\end{abstract}

\maketitle

\section{Introduction}

The properties of hadrons are of great interest to understand the
nonperturbative behavior of the fundamental theory of strong
interactions, quantum chromodynamics (QCD).  Dynamical quark-mass
generation is a subject to be studied, where chiral symmetry plays a
key role.  An order parameter of the spontaneous breakdown of chiral
symmetry in the QCD vacuum is the chiral condensate.  The masses of
the light vector mesons ($\rho$, $\omega$, and $\phi$) are considered
to be mostly induced by this order parameter.  In this regard, mass
modifications of the vector mesons at finite density and/or finite
temperature have been studied both theoretically and
experimentally~\cite{Metag:2017yuh}.  No clear evidence for them has
been observed so far.  Another candidate to study the relationship
between the mass and chiral condensate is the $\eta ^{\prime} (958)$
meson.  It has an exceptionally large mass although it would be a
Nambu-Goldstone boson originating from the $\text{U}_{\rm L}(3) \times
\text{U}_{\rm R}(3)$ chiral symmetry breaking~\cite{Weinberg:1975ui}.
Its mass generation is considered to be a result of the quantum
anomaly in QCD which breaks $\text{U}_{\rm A} ( 1 )$
symmetry~\cite{tHooft:1976rip, Witten:1979vv, Veneziano:1979ec}.  In
addition, it was also pointed out that the chiral condensate plays an
essential role for the anomaly to affect the $\eprime$
mass~\cite{Cohen:1996ng, Lee:1996zy}.

In this line, various studies on the in-medium properties of the
$\eprime$ meson are performed to understand QCD in the nuclear medium
theoretically~\cite{Jido:2011pq, Pisarski:1983ms, Bernard:1987sg,
  Kunihiro:1989my, Kapusta:1995ww, Tsushima:1998qp, Costa:2002gk,
  Nagahiro:2004qz, Bass:2005hn, Nagahiro:2006dr, Nagahiro:2011fi,
  Sakai:2013nba, Bass:2013nya} and experimentally~\cite{Nanova:2012vw,
  Nanova:2013fxl, Nanova:2016cyn, Tanaka:2016bcp}.  In particular,
from the theoretical side, assuming the mass difference between $\eta
^{\prime}$ and low-lying pseudoscalar mesons comes from the chiral
condensate in connection with the $\text{U}_{\rm A}(1)$ anomaly, we
expect that the $\eprime$ mass will be reduced by an order of 100 MeV
at normal nuclear density, because partial restoration of chiral
symmetry in a nuclear medium, which was suggested by pionic atoms as a
reduction of the chiral order parameter~\cite{Suzuki:2002ae}, induces
suppression of the $\text{U}_{\rm A}(1)$ anomaly effect to the
$\eprime$ mass~\cite{Jido:2011pq}.  A chiral effective model
calculation by the linear $\sigma$ model implies a $\eprime$ mass
reduction of $\sim 80 \mev$ at normal nuclear
density~\cite{Sakai:2013nba}.  A more sophisticated calculation based
on the Nambu--Jona-Lasinio model, in which the $\text{U}_{\rm A}(1)$
effect is introduced by the Kobayashi-Maskawa-'t~Hooft term, predicts
a large reduction of approximately 150 MeV at normal nuclear
density~\cite{Costa:2002gk}.  Such a large reduction of the $\eprime$
mass allows formation of $\eta ^{\prime}$-nucleus bound states ($\eta
^{\prime}$-mesic nuclei).  From the experimental side, the results
obtained by the CBELSA/TAPS collaboration imply an attractive and
weakly absorptive potential~\cite{Metag:2017yuh, Nanova:2012vw,
  Nanova:2013fxl, Nanova:2016cyn}: the real part of the
$\eprime$-nucleus potential at normal nuclear density was found to be
$- 37 \pm 10 (\text{stat}) \pm 10 (\text{syst}) \mev$ in the $\eprime$
photoproduction from ${}^{12}\text{C}$~\cite{Nanova:2013fxl} and $- 41
\pm 10 (\text{stat}) \pm 15 (\text{syst}) \mev$ from
${}^{93}\text{Nb}$~\cite{Nanova:2016cyn}, while its imaginary part was
found to be $- ( 10 \pm 2.5 ) \mev$~\cite{Nanova:2012vw}.  At GSI, the
excitation spectrum for the ${}^{12}{\rm C}(p,d)$ reaction was
measured to search for $\eta ^{\prime}$-mesic
nuclei~\cite{Tanaka:2016bcp}.  The result of the GSI experiment seems
to exclude strongly attractive $\eta ^{\prime}$-nucleus potential with
a mass reduction of $\gtrsim 150 \mev$ at normal nuclear density, but
is still consistent with the $\eprime$ mass reduction of $\lesssim 80
\mev$.  To pin down the properties of the $\eprime$ meson in nuclei
more rigorously, we need various experimental information on
$\eprime$-nucleon and $\eprime$-nucleus systems, such as the $\eprime
N$ scattering length in free space~\cite{Czerwinski:2014yot}.

We here emphasize that the $\eprime$ mass reduction in a nuclear
medium is induced by an attractive $\eprime N$ interaction.  In this
sense, the $\eprime N$ interaction plays a key role to investigate
properties of the $\eprime$ meson.  Because experimental information
on the $\eprime N$ interaction is not sufficient, we employ symmetry
properties of hadrons to deduce the $\eprime N$ interaction.  The
$\eprime N$ interaction was studied in, e.g., the chiral effective
model~\cite{Kawarabayashi:1980uh, Bass:1999is, Borasoy:1999nd,
  Oset:2010ub, Sakai:2013nba}.  In terms of the linear $\sigma$ model,
the scalar meson exchange provides an attractive interaction between
$\eta ^{\prime}$ and nucleon which is strong enough to bind the
$\eprime N$ system~\cite{Sakai:2013nba}.  Experimentally, the
existence of an $\eprime N$ bound state is implied by near-threshold
behavior of the total cross section of the $\pi ^{-} p \to \eprime n$
reaction~\cite{Moyssides:1983}.  It has been also pointed out that
$\eprime n$ bound state, if exists, can be observed in incoherent
photoproduction off a deuteron target $\gamma d \to \eta n
p$~\cite{Sekihara:2016jgx}.

In this study we extend the consideration on the $\eprime N$ system to
the $\eprime d$ system.  We take the $\eprime N$ interaction from the
linear $\sigma$ model~\cite{Sakai:2013nba} and solve the Faddeev
equation for the $\eprime d$ system in a certain approximation.  We
will see that the $\eprime d$ system is bound thanks to a strongly
attractive $\eprime N$ interaction in our model.  We further discuss
whether this $\eprime d$ bound state can be observed in experiments or
not.  For this purpose we choose the $\gamma d \to \eta d$ reaction,
in which the $\eprime d$ bound state can be formed in the $s$-channel
process and decays into $\eta d$.  The biggest advantage of this
reaction is that we can easily perform the center-of-mass energy scan
to search for the $\eprime d$ bound state by varying the photon
energy.  Because the $\eprime d$ threshold is $2.833 \gev$, the photon
energy appropriate for the bound-state search is around $ 1.20 \gev$
in the laboratory frame.  In addition, it is worth mentioning that the
final-state $\eta d$ can specify an isospin $0$ state in its $s$
channel.  In the following we will formulate the $\gamma d \to \eta d$
reaction mechanism and calculate its cross section to estimate the
production cross section of the $\eprime d$ bound state.

This paper is organized as follows.  In Sec.~\ref{sec:2} we show
that the $\eprime N$ interaction from the linear $\sigma$
model leads to an $\eprime d$ bound state.  Next, in Sec.~\ref{sec:3}
we evaluate the cross section of the $\gamma d \to \eta d$ reaction,
in which an $\eprime d$ bound state may be generated, by using
phenomenological $\gamma N \to \eboth N$ amplitudes and the $\eprime
N$ interaction constructed in the linear $\sigma$ model.
Section~\ref{sec:4} is devoted to the summary of this paper.
Throughout this study we assume isospin symmetry for hadron masses as
well as strong interactions.

\section{\boldmath Possible $\eprime d$ bound state}
\label{sec:2}

\subsection{\boldmath $\eprime N$ system}
\label{sec:2A}

\begin{figure}[!t]
  \centering
  \Psfig{8.6cm}{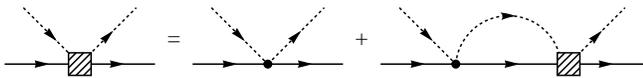} 
  \caption{Diagrammatic equation for the $\eboth N \to \eboth N$
    scattering amplitude.  The solid and dashed lines represent the
    nucleon and $\eboth$, respectively.  The shaded squares and dots
    are the full scattering amplitude and tree-level interaction,
    respectively.}
  \label{fig:T}
\end{figure}

First of all, we consider the $\eprime N$ interaction.  We focus on
the $s$-wave $\eprime N$ system, and take into account a coupling to
the $\eta N$ channel because it is the closest open channel coupled in
the $s$ wave.  We employ the $\eprime N$ interaction in the linear
$\sigma$ model with the unitarization according to
Ref.~\cite{Sakai:2013nba}.  Dynamics in the $\eprime p$ and $\eprime
n$ systems is the same, because we assume isospin symmetry.  We assign
a channel index of $1$ ($2$) to the $\eprime N$ ($\eta N$) channel.
In the linear $\sigma$ model, the $\eprime N$ interaction can be
described by the exchange of the singlet and octet $\sigma$ mesons.
In momentum space, the interaction $V_{j k}$ with the channel indices
$j$ and $k$ can be written as~\cite{Sakai:2013nba}
\begin{equation}
  V_{1 1} = - \frac{6 g B}{\sqrt{3} m_{\sigma _{0}}^{2}} ,
  \quad 
  V_{1 2} = V_{2 1} = + \frac{6 g B}{\sqrt{6} m_{\sigma _{8}}^{2}} ,
  \quad 
  V_{2 2} = 0 ,
  \label{eq:Vjk}
\end{equation}
where the constants $g$, $B$, $m_{\sigma _{0}}$, and $m_{\sigma _{8}}$
determine the strength of the interaction; $g$ is the $\sigma N N$
coupling constant, $B$ is the contribution from the $\text{U}_{\rm A}
( 1 )$ anomaly, and $m_{\sigma _{0}}$ and $m_{\sigma _{8}}$ are the
masses of the singlet and octet $\sigma$ mesons.

It should be noted that the interaction in Eq.~\eqref{eq:Vjk} is the
leading-order term of the momentum expansion in the flavor SU(3)
symmetric limit.  When we switch on the flavor symmetry breaking by
the heavier strange quark than up and down quarks, the
$\eta$-$\eprime$ mixing angle is $-6.2$ degrees in the linear $\sigma$
model~\cite{Sakai:2013nba}.  Even in this case of the linear $\sigma$
model, the modification of the $\eprime N$ interaction by the
$\eta$-$\eprime$ mixing is small~\cite{Sakai:2013nba}, owing to the
dominance of the singlet $\eta$ in the physical $\eprime$ state.
Furthermore, the strength of the $\eprime N \to \eprime N$ part, which
is crucial in the following discussions, shifts only several percent
for the $\eta$-$\eprime$ mixing angle between $0$ and $-20$ degrees,
which is adopted in Ref.~\cite{Bass:2013nya}.

Then the $\eprime N$ scattering amplitude $T_{j k} ( w )$, as a
function of the energy of the $\eprime N$ system $w$, is a solution of
the Lippmann--Schwinger equation diagrammatically expressed in
Fig.~\ref{fig:T}.  This equation can be written in the present
formulation as
\begin{equation}
T_{j k} ( w ) 
= V_{j k} + \sum _{l = 1}^{2} V_{j l} G_{l} ( w ) T_{l k} ( w ) 
\label{eq:LSeq}
\end{equation}
with the $\eboth N$ loop function $G_{j}$.  Because the interaction
$V_{j k}$ is independent of the external momentum as in
Eq.~\eqref{eq:Vjk}, the scattering equation~\eqref{eq:LSeq} becomes 
algebraic.  For the $\eboth N$ loop function, we employ a
covariant expression as
\begin{equation}
  G_{j} ( w ) \equiv 
  i \int \frac{d^{4} q}{(2 \pi )^{4}} 
  \frac{2 m_{N}} {[(p - q)^{2} - m_{N}^{2} + i 0] (q^{2} - m_{j}^{2}  + i 0)} 
\end{equation}
with $p^{\mu} = (w , \, \bm{0})$, the nucleon mass $m_{N}$, and $m_{1}
= m_{\eprime}$ and $m_{2} = m_{\eta}$ being the $\eprime$ and $\eta$
masses, respectively.  The loop function is calculated with the
dimensional regularization as
\begin{align}
& G_{j} (w) = \frac{2 m_{N}}{16 \pi ^{2}} \Bigg [ a_{j} ( \mu _{\rm reg} ) 
+ \ln \left ( \frac{m_{N}^{2}}{\mu _{\rm reg}^{2}} \right )  
\notag \\
& + \frac{w^{2} + m_{j}^{2} - m_{N}^{2}}{2 w^{2}} 
\ln \left ( \frac{m_{j}^{2}}{m_{N}^{2}} \right ) 
\notag \\
& - \frac{\lambda ^{1/2} (w^{2}, \, m_{N}^{2}, \, m_{j}^{2})}{w^{2}} 
\text{arctanh} 
\left ( \frac{\lambda ^{1/2} (w^{2}, \, m_{N}^{2}, \, m_{j}^{2})}
{m_{N}^{2} + m_{j}^{2} - w^{2}} \right ) 
\Bigg ] 
\label{eq:Gdim_explicit}
\end{align}
with the regularization scale $\mu _{\rm reg}$, the subtraction
constant $a_{j}$, and $\lambda ( x , \, y , \, z ) \equiv x^{2} +
y^{2} + z^{2} - 2 x y - 2 y z - 2 z x$.  In this study the subtraction
constant is fixed by the natural renormalization scheme developed in
Ref.~\cite{Hyodo:2008xr} so as to exclude the Castillejo-Dalitz-Dyson
pole contribution from the loop function.  This can be achieved by
requiring $G_{j} ( w = m_{N} ) = 0$ for every channel $j$, which
results in $a_{1} ( \mu _{\rm reg} = m_{N} ) = -1.84$ and $a_{2} ( \mu
_{\rm reg} = m_{N} ) = -1.24$ in the present construction.

Now we fix the model parameters as in Ref.~\cite{Sakai:2013nba}, i.e.,
$g = 7.67$, $B = 0.984 \gev$, $m_{\sigma _{0}} = 0.7 \gev$, and
$m_{\sigma _{8}} = 1.23 \gev$, with which we obtain an $\eprime N$
bound state.  The pole position of the $\eprime N$ bound state is
$1889 - 6 i \mev$, which corresponds to the binding energy of $8 \mev$
measured from the $\eprime N$ threshold and decay width of $12 \mev$.
The existence of an $\eprime N$ bound state is implied by
near-threshold behavior of the total cross section of the $\pi ^{-} p
\to \eprime n$ reaction~\cite{Moyssides:1983}, although no bump
corresponding to the $\eprime n$ bound state has been observed in the
$\gamma d \to p X$ reaction at LEPS~\cite{Muramatsu}.  In the $\eprime
N$ scattering, the contribution from the $\eta N$ channel is found to
be small while the elastic $\eprime N$ interaction is dominant.  This
is because the transition of the $\eprime N$ channel to the $\eta N$
channel is suppressed by the larger mass of the octet scalar meson.

\subsection{\boldmath $\eprime d$ system}
\label{sec:2B}

Next, using the $\eprime N$ interaction constructed in the previous
subsection, we formulate the $\eprime d$ scattering amplitude.  We
treat the $\eprime p n$ three-body system, where we consider the $p n$
subsystem as a deuteron and solve the Faddeev equation in the
so-called fixed center approximation (FCA)~\cite{Bayar:2011qj,
  Sekihara:2016vyd}.  We incorporate two channels for the three-body
system: 1) $\eprime p n$ and 2) $p n \eprime$.  We distinguish either
$\eprime$ appears in the left or right, according to the formulation
in Ref.~\cite{Sekihara:2016vyd}.  For instance, if the initial state
is $\eprime p n$ ($p n \eprime$), the multiple scattering starts with
the $\eprime p$ ($\eprime n$) scattering in the system.  Similarly, if
the final state is $\eprime p n$ ($p n \eprime$), the multiple
scattering ends with the $\eprime p$ ($\eprime n$) scattering in the
system.  Besides, we can fix the ordering of the nucleons, $p n$,
without loss of generality.  In the three-body problem, the $\eta$
meson does not appear explicitly but is intrinsically treated in the
two-body $\eprime N \to \eprime N$ amplitude.

\begin{figure}[!t]
  \centering
  \Psfig{8.6cm}{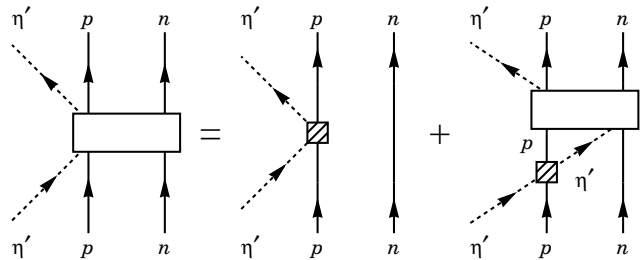}
  \caption{Diagrammatic equation for the multiple $\eprime$ scattering
    amplitude of the process $\eprime p n \to \eprime p n$.  The small
    shaded boxes represent the $\eprime N \to \eprime N$ scattering
    amplitude, and the large open boxes indicate its multiple
    scattering amplitude.}
  \label{fig:FCA}
\end{figure}

In order to grasp the construction, we first consider the $\eprime p n
\to \eprime p n$, i.e., channel $1 \to 1$ scattering amplitude
$T_{11}^{\rm FCA}$.  This is schematically expressed in
Fig.~\ref{fig:FCA} as a diagrammatic equation and can be written as
\begin{align}
  T^{\rm FCA}_{1 1} ( W ) = & t_{1} ( W ) 
  + t_{1} ( W ) G_{\eprime}^{\rm FCA} ( W ) T^{\rm FCA}_{2 1} ( W ) 
  \label{eq:TFCA_11}
\end{align}
with the total three-body energy $W$ and the three-body Green function
$G_{\eprime}^{\rm FCA}$ of the $\eprime$ propagation.  The two-body
$\eprime N \to \eprime N$ scattering amplitude $t_{1}$ is developed in
the previous subsection
\begin{equation}
  t_{1} ( W ) = T_{1 1} ( w^{\rm FCA} ( W ) ) 
\end{equation}
with the $\eprime N$ two-body center-of-mass energy $w^{\rm FCA}$.  We
evaluate the two-body energy $w^{\rm FCA}$ as a function of the
three-body energy $W$~\cite{Bayar:2011qj, Sekihara:2016vyd} by
treating two nucleons as one particle of mass $2 m_{N}$:
\begin{equation}
  w^{\rm FCA} ( W ) =
  \sqrt{\frac{W^{2} + m_{\eprime}^{2} - 2 m_{N}^{2}}{2}} .
  \label{eq:wFCA}
\end{equation}
The three-body Green function $G_{\eprime}^{\rm FCA}$ is defined as:
\begin{equation}
  G_{\eprime}^{\rm FCA} ( W ) = \int \frac{d^{3} p}{( 2 \pi )^{3}}
  \frac{F_{N N} ( p )}{p_{\eprime}^{0} ( W )^{2} - \bm{p}^{2} 
    - m_{\eprime}^{2} + i 0}
\end{equation}
with the $\eprime$ energy $p_{\eprime}^{0}$
\begin{equation}
  p_{\eprime}^{0} ( W ) = \frac{W^{2} + m_{\eprime}^{2} - (2 m_{N})^{2}}{2 W} ,
  \label{eq:peta0}
\end{equation}
and the deuteron form factor $F_{N N} ( p )$
\begin{equation}
  F_{N N} ( p ) = \int d^{3} r e^{i \bm{p} \cdot \bm{r}}
  | \varphi ( r ) |^{2} .
\end{equation}
Here $\varphi ( r )$ is the deuteron wave function in coordinate
space, and the form factor can be rewritten as
\begin{equation}
  F_{N N} ( p ) = \int \frac{d^{3} q}{( 2 \pi )^{3}}
  \deutwf ( q ) \deutwf ( | \bm{q} - \bm{p} | ) 
  \label{eq:F_NN}
\end{equation}
with the deuteron wave function in momentum space $\deutwf ( q )$.
For the deuteron wave function, we neglect the $d$-wave component and
use a parameterization of the $s$-wave component given in an analytic
function~\cite{Lacombe:1981eg} as
\begin{equation} 
\deutwf ( q ) =
\sum _{j=1}^{11} \frac{C_{j}}{q^{2} + m_{j}^{2}} 
\label{eq:deutWF}
\end{equation}
with $C_{j}$ and $m_{j}$ determined with the charge-dependent Bonn
potential~\cite{Machleidt:2000ge}.  This wave function is normalized
so as to satisfy $F_{N N} ( p = 0 ) = 1$.

\begin{figure}[!t]
  \centering \Psfig{8.6cm}{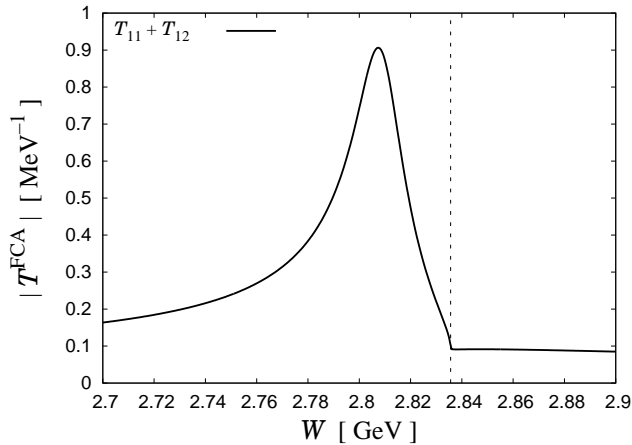}
  \caption{Absolute value of the scattering amplitude $T_{1 1}^{\rm
      FCA} + T_{1 2}^{\rm FCA}$ as a function of the total three-body
    energy $W$.  The vertical dotted line indicates the $\eprime N N$
    threshold.}
  \label{fig:FCAabs}
\end{figure}

\begin{figure*}[!t]
  \centering
  \PsfigII{0.190}{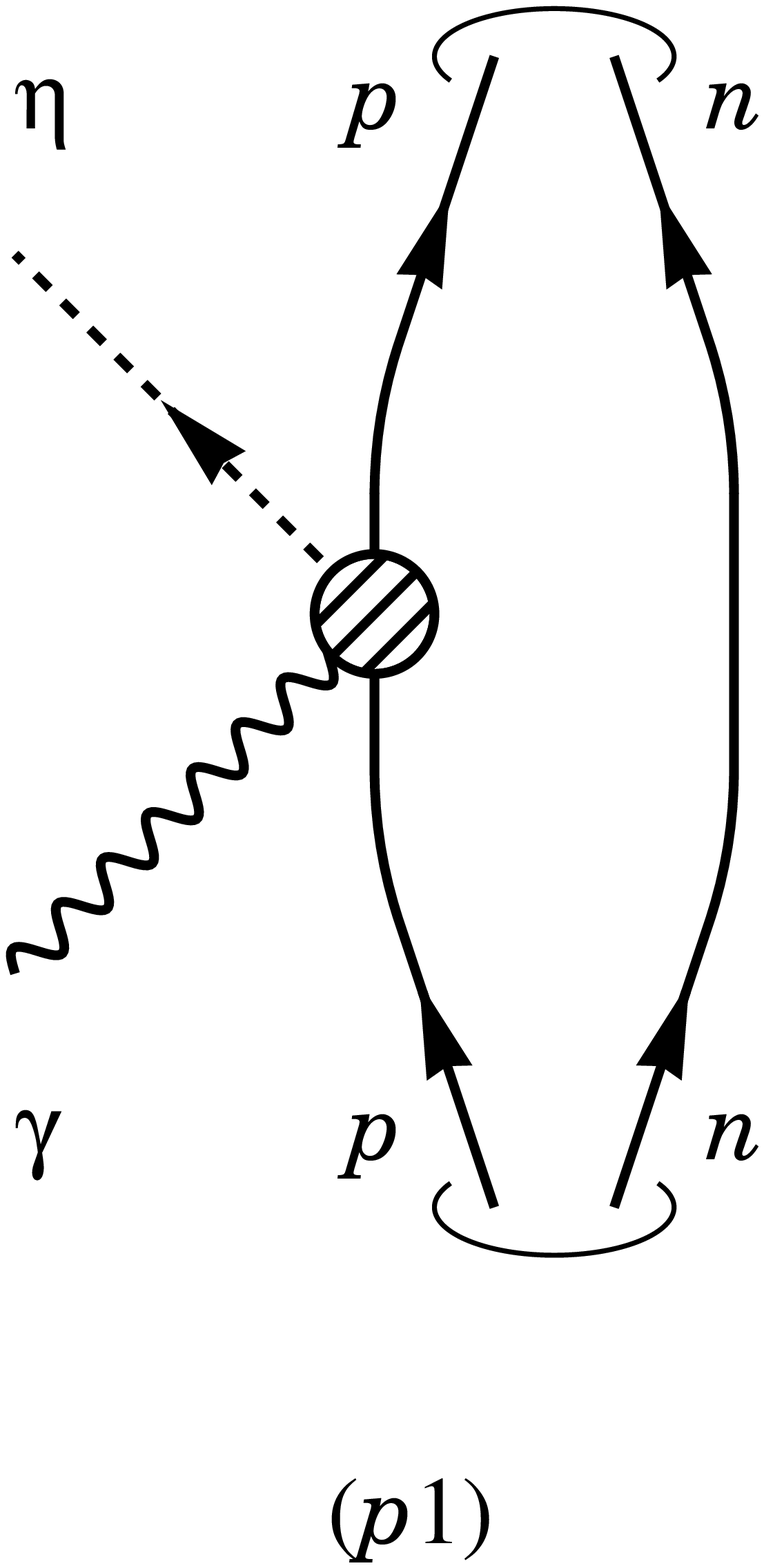} ~ 
  \PsfigII{0.190}{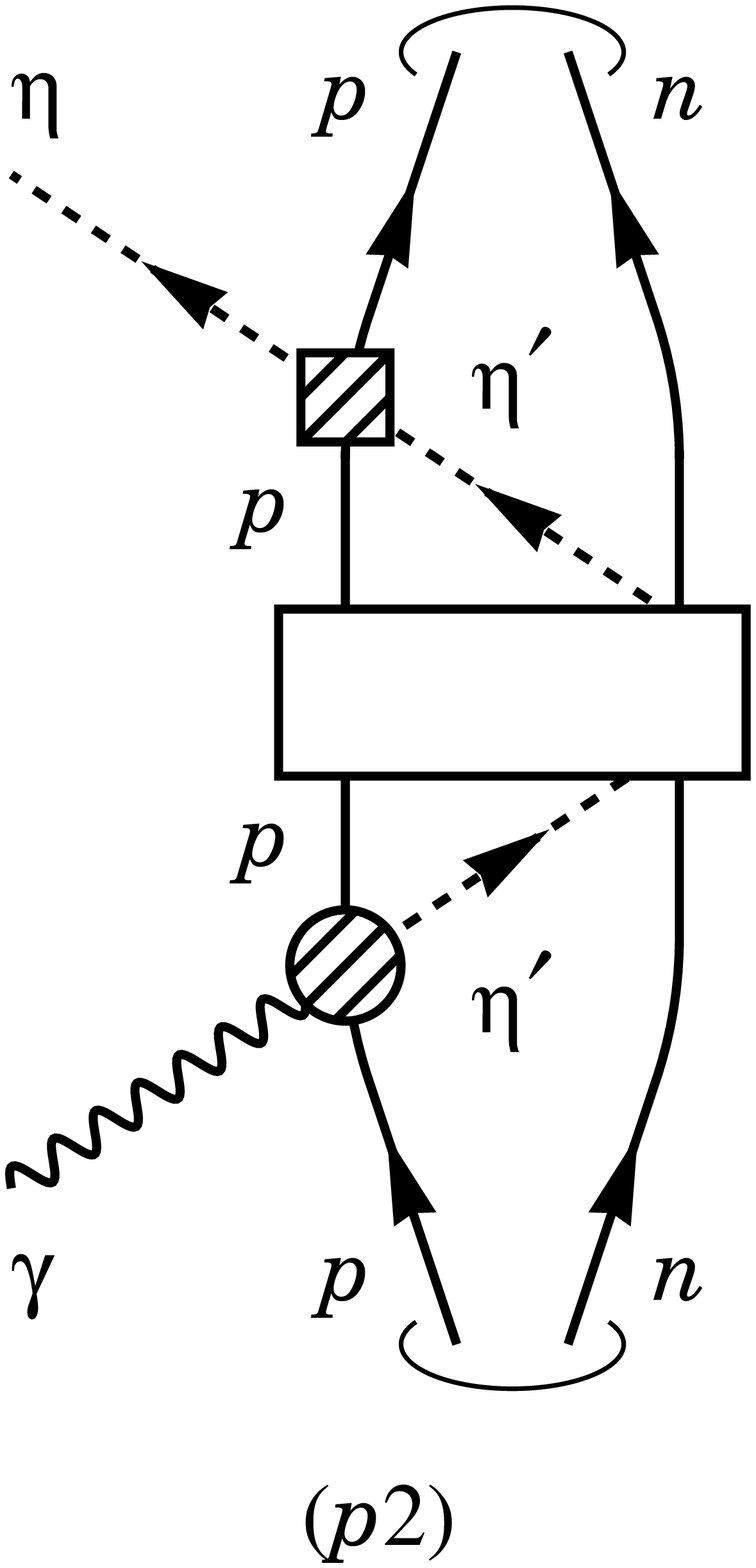} ~ 
  \PsfigII{0.190}{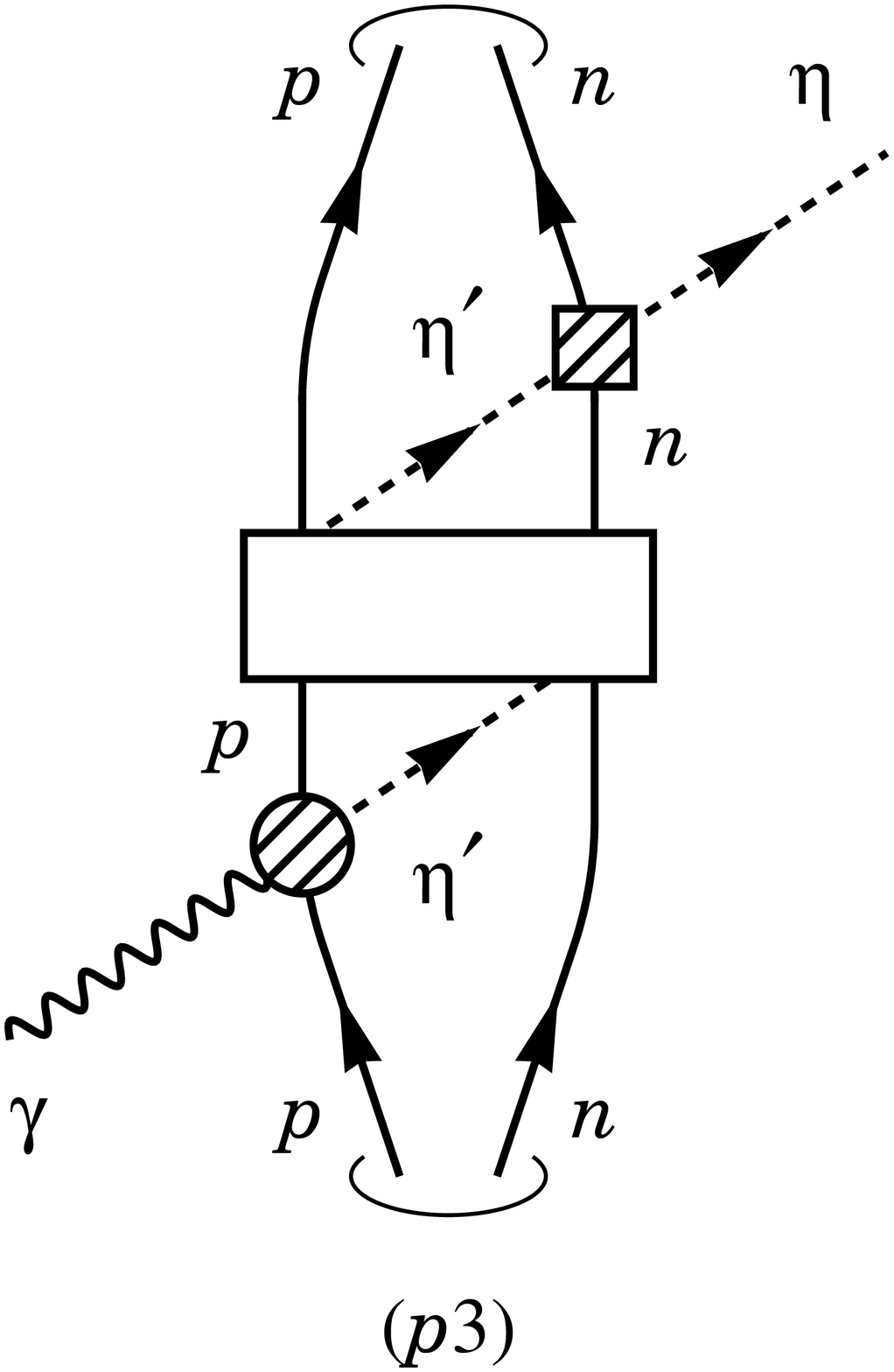} ~ 
  \PsfigII{0.190}{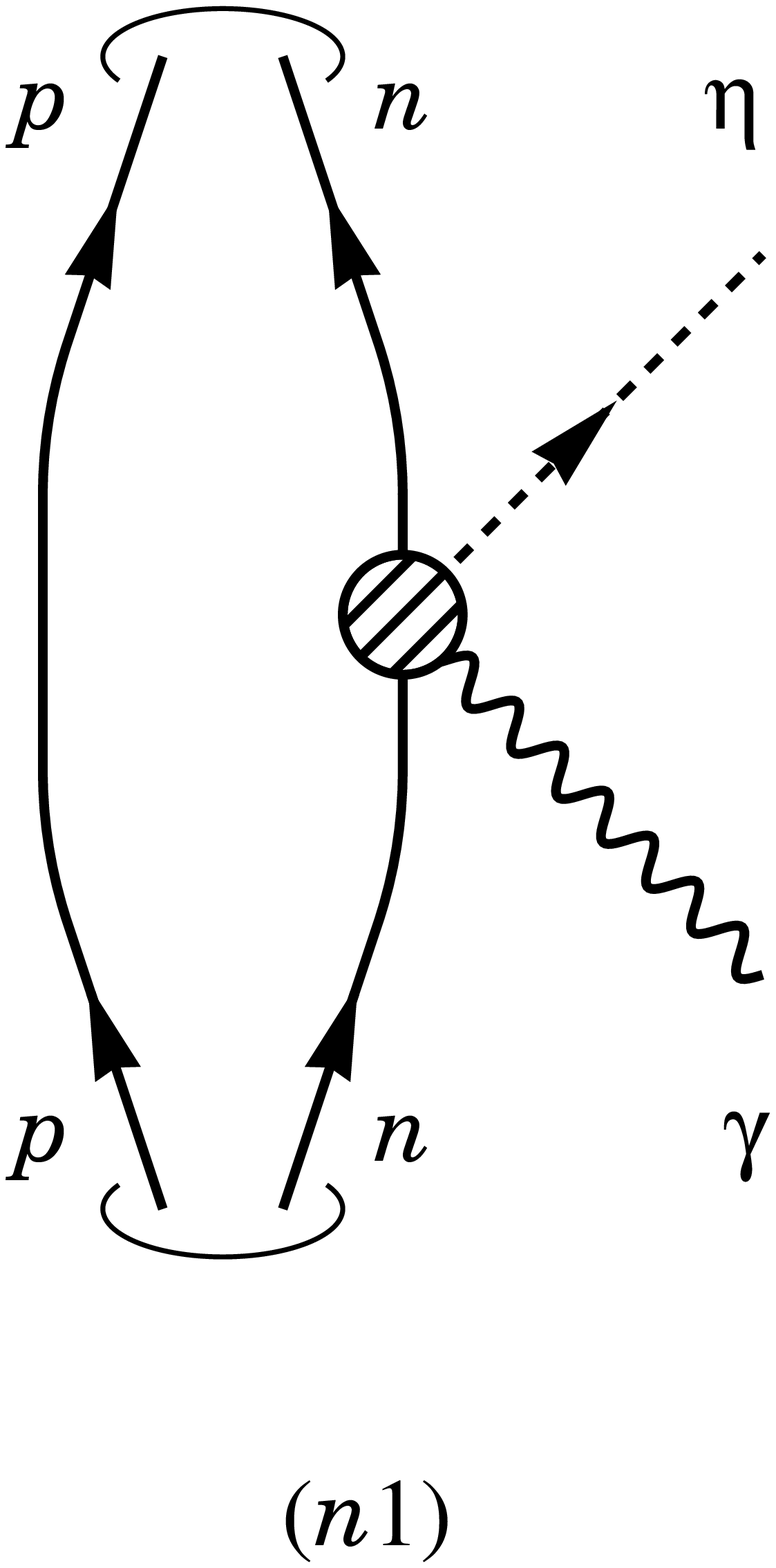} ~ 
  \PsfigII{0.190}{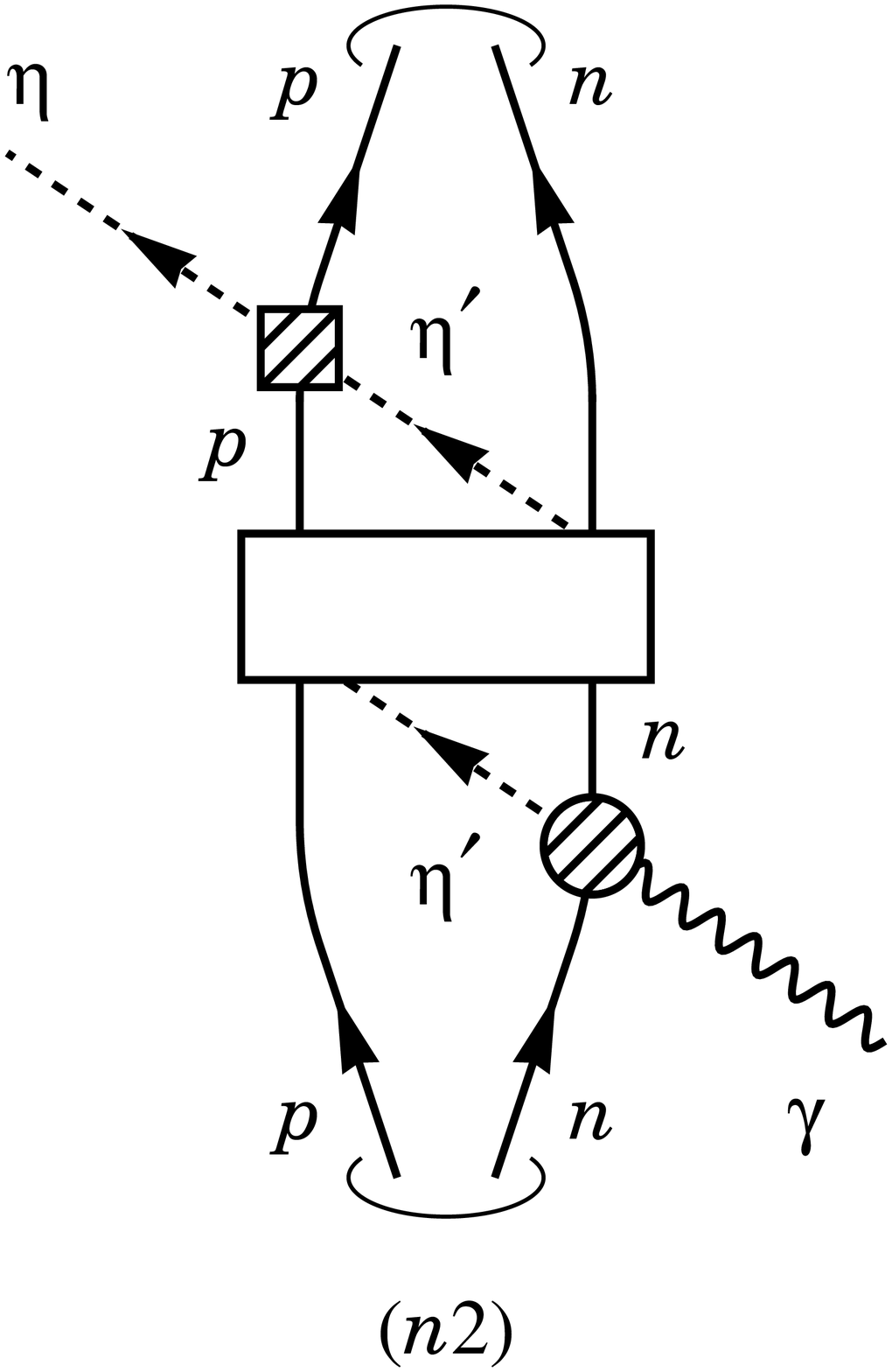} ~
  \PsfigII{0.190}{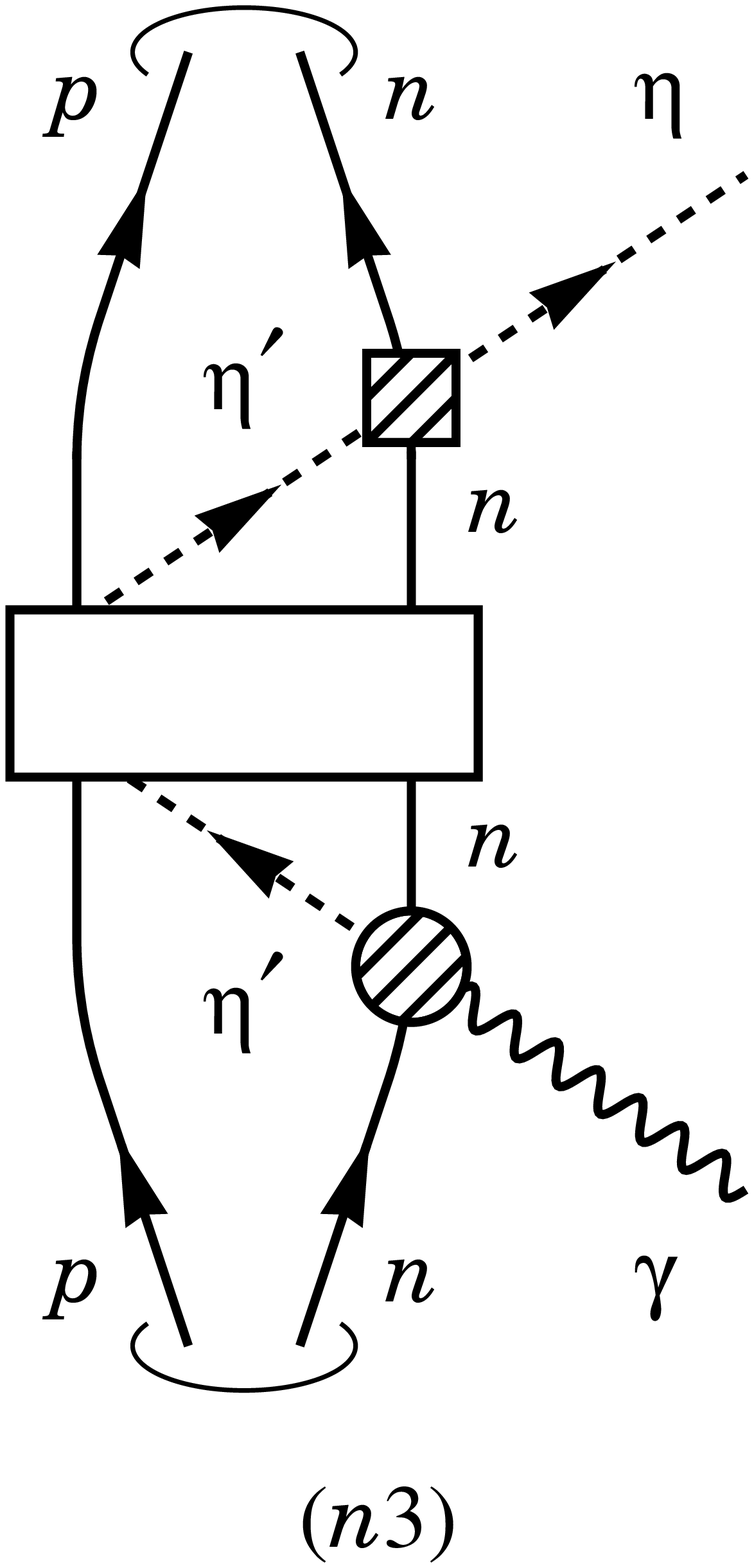} 
  \caption{Diagrams for the $\gamma d \to \eta d$ reaction.  The
    shaded circles represent the $\gamma N \to \eboth N$ amplitude.
    The small shaded boxes indicate the $\eprime N \to \eta N$
    amplitude.  The large open boxes represent the multiple scattering
    amplitude for the $\eprime p n$ system.}
  \label{fig:diag_coh}
\end{figure*}

The scattering equation~\eqref{eq:TFCA_11} can be straightforwardly
extended to the two-channel case, and we obtain
\begin{align}
  & T^{\rm FCA}_{a b} ( W ) 
  = V_{a b}^{\rm FCA} ( W )
  \notag \\
  & \quad + \sum _{c = 1}^{2} \tilde{V}_{a c}^{\rm FCA} ( W ) G_{c}^{\rm FCA} ( W ) 
  T^{\rm FCA}_{c b} ( W ) .
\end{align}
Here $a$, $b$, and $c$ ($=1$, $2$) are three-body channel indices and
$V_{a b}^{\rm FCA}$ and $\tilde{V}_{a b}^{\rm FCA}$ contain the
$\eprime N \to \eprime N$ scattering amplitude as follows:
\begin{equation}
  V_{a b}^{\rm FCA} = \left (
  \begin{array}{@{\,}cc@{\,}}
    t_{1} & 0 \\
    0 & t_{1} \\
  \end{array}
  \right ) ,
  \quad
  \tilde{V}_{a b}^{\rm FCA} = \left (
  \begin{array}{@{\,}cc@{\,}}
    0 & t_{1} \\
    t_{1} & 0 \\
  \end{array}
  \right ) .
\end{equation}
The three-body loop function $G_{a}^{\rm FCA}$ is
\begin{equation}
  G_{1}^{\rm FCA} = G_{2}^{\rm FCA} = G_{\eprime}^{\rm FCA} .
\end{equation}
With this formulation, we can calculate the $\eprime d$ scattering
amplitude as a function of the total three-body energy $W$.

As we will see later, the multiple scattering amplitude practically
appears as the sum of the $\eprime p n$ and $p n \eprime$
contributions, such as $T_{1 1}^{\rm FCA} + T_{1 2}^{\rm FCA}$, in
full reaction amplitudes.  The absolute value of this scattering
amplitude $T_{1 1}^{\rm FCA} + T_{1 2}^{\rm FCA}$ is shown in
Fig.~\ref{fig:FCAabs} as a function of the total three-body energy
$W$.  As one can see, the amplitude has a peak structure corresponding
to the $\eprime d$ bound state.  In the amplitude we find the pole of
the $\eprime d$ bound state at $2809 - 10 i \mev$ in the complex
energy plane, which corresponds to the binding energy $25 \mev$
measured from the $\eprime d$ threshold and width $19 \mev$.  The
binding energy increases more than twice compared to the $\eprime N$
bound state because the number of the potential terms increases more
than that of the kinetic energy, as in usual many-body systems.

\section{\boldmath $s$-channel formation of the $\eprime d$ bound
  state in the $\gamma d \to \eta d$ reaction}
\label{sec:3}

Because the $\eprime d$ system is bound with the $\eprime N$
interaction deduced from the linear $\sigma$ model, it may be
experimentally generated in certain reactions.  In this section, we
consider the $\gamma d \to \eta d$ reaction and examine a possibility
of observing its signal.  We first formulate the $\gamma d \to \eta d$
scattering amplitude in Sec.~\ref{sec:3A}, and show the numerical
results in Sec.~\ref{sec:3B}.

\subsection{\boldmath Formulation}
\label{sec:3A}

In order to calculate the scattering amplitude of the $\gamma d \to
\eta d$ reaction, we introduce six diagrams relevant to the formation
of the $\eprime d$ bound state as shown in Fig.~\ref{fig:diag_coh}:
\begin{align}
  T_{\gamma d \to \eta d} = & \mathcal{T}_{p 1} + \mathcal{T}_{p 2} 
  + \mathcal{T}_{p 3} + \mathcal{T}_{n 1} + \mathcal{T}_{n 2} 
  + \mathcal{T}_{n 3} .
  \label{eq:Tgammad}
\end{align}
On the one hand, $\mathcal{T}_{p 1}$ and $\mathcal{T}_{n 1}$ are the
single-step scatterings for the reaction, which becomes a background
in view of the signal of the $\eprime d$ bound-state formation.  On
the other hand, the remaining four terms contain the multiple
$\eprime$ scattering on both $p$ and $n$ which generates the $\eprime
d$ bound state.  We here neglect diagrams in which the $\eta$ meson is
produced in the intermediate state, because around the $\eprime d$
threshold the $\eta$ meson in the intermediate state should go highly
off-shell and should be kinematically suppressed.  This resembles the
case of photoproduction of the $\eprime n$ bound state in the $\gamma
d \to \eta n p$ reaction, as discussed in
Ref.~\cite{Sekihara:2016jgx}.  The reaction diagrams in
Fig.~\ref{fig:diag_coh} contain the $\gamma N \to \eta N$ and $\gamma
N \to \eprime N$ scattering amplitudes and the transition amplitude of
the $\eprime d$ bound state to the final-state $\eta d$ system.

Below, we formulate the $\gamma N \to \eta N$ and $\gamma N \to
\eprime N$ scattering amplitudes based on the experimental data.  We
then construct the $\gamma d \to \eta d$ scattering
amplitude~\eqref{eq:Tgammad} from the amplitudes of $\gamma N \to
\eboth N$, multiple $\eprime$ scattering on $p n$, and transition to
$\eta d$.  In the present formulation of the $\gamma d \to \eta d$
amplitude, we will fix the photo-induced $\gamma N \to \eta ^{( \prime
  )} N$ amplitudes so as to reproduce the existing experimental data
of $\eta ^{( \prime )}$ photoproduction.  Therefore, when we modify
the $\eprime N$ interaction, they affect only the amplitudes of the
multiple $\eprime$ scattering on $p n$ and of $\eprime N \to \eta N$
entering in the $\eprime d \to \eta d$ transition in our model.

\subsubsection{$\gamma N \to \eta N$ and $\gamma N \to \eprime N$
  scattering amplitudes}

Let us consider the $\gamma p \to \eta p$ and $\gamma n \to \eta n$
scattering amplitudes.  For these reactions, there exist various
experimental data of the differential cross sections as a function of
the photon energy in the laboratory frame $\Eglab$ and the $\eta$
scattering angle in the center-of-mass frame $\theta _{\eta}$, around
the photon energy of interest, $\Eglab \approx 1.2 \gev$: for
instance, the free proton target case~\cite{Nakabayashi:2006ut,
  Bartholomy:2007zz, Williams:2009yj, Sumihama:2009gf, Crede:2009zzb,
  McNicoll:2010qk} and the deuteron target case~\cite{Jaegle:2011sw,
  Werthmuller:2014thb, Ishikawa:2016rgk, Witthauer:2017pcy}.  Several
theoretical analyses of these data are available as well, e.g., in
Refs.~\cite{Chiang:2002vq, Anisovich:2011fc, Kamano:2013iva,
  Ronchen:2015vfa}.

For the $\gamma p \to \eta p$ reaction, we take the theoretical values
of the differential cross section summarized by the Bonn--Gatchina
partial wave analysis (BG2014-02)~\cite{Gutz:2014wit}.  We simply
translate these values into the scattering amplitudes as functions of
$\Eglab$ and $\theta _{\eta}$ through the formula:
\begin{equation}
  T_{\gamma p \to \eta p} ( \Eglab , \, \cos \theta _{\eta} )
  = \sqrt{\frac{16 \pi ^{2} q_{\rm cm} w^{2}}{m_{N}^{2} q_{\rm cm}^{\prime}}
    \frac{d \sigma _{\gamma p \to \eta p}}{d \Omega}} ,
  \label{eq:TgNetaN}
\end{equation}
for the $\gamma p \to \eta p$ reaction.  Here $w$ is the
center-of-mass energy and $q_{\rm cm}$ and $q_{\rm cm}^{\prime}$ are
the relative momenta of the initial- and final-state particles in the
center-of-mass frame, respectively.  For the later convenience, we
show the explicit form of $w$, $q_{\rm cm}$, and $q_{\rm cm}^{\prime}$
as functions of $\Eglab$:
\begin{equation}
  w ( \Eglab ) = \sqrt{m_{N}^{2} + 2 m_{N} \Eglab },
  \label{eq:w-Eg}
\end{equation}
\begin{equation}
  q_{\rm cm} ( \Eglab ) = \frac{w ( \Eglab ) ^{2} - m_{N}^{2}}{2 w ( \Eglab )} ,
  \label{eq:qcm}
\end{equation}
and
\begin{equation}
  q_{\rm cm}^{\prime} ( \Eglab ) = 
    \frac{\lambda ^{1/2} ( w ( \Eglab )^{2}, \, m_{\eta}^{2}, \, m_{N}^{2} )}
    {2 w ( \Eglab )} .
  \label{eq:qcmp}
\end{equation}

As for the $\gamma n \to \eta n$ amplitude $T_{\gamma n \to \eta n}$,
one could evaluate it in a similar manner, but here we recall a
general relation for $\eta$ photoproduction:
\begin{equation}
  T_{\gamma p \to \eta p} \propto A_{\rm IS} + A_{\rm IV} , 
  \quad 
  T_{\gamma n \to \eta n} \propto A_{\rm IS} - A_{\rm IV} , 
\end{equation}
where $A_{\rm IS}$ denotes the isoscalar amplitude and $A_{\rm IV}$
the isovector one.  In coherent $\eta$ photoproduction off the
deuteron, only the sum $T_{\gamma p \to \eta p} + T_{\gamma n \to \eta
  n} \propto 2 A_{\rm IS}$ contributes to the full amplitude.
Therefore, we may write the sum of the amplitude as
\begin{align}
  & T_{\gamma p \to \eta p} ( \Eglab , \, \cos \theta _{\eta} )
  + T_{\gamma n \to \eta n} ( \Eglab , \, \cos \theta _{\eta} )
  \notag \\
  & = \frac{2 | A_{\rm IS} |}{| A_{\rm IS} + A_{\rm IV} |}
  \sqrt{\frac{16 \pi ^{2} q_{\rm cm} w^{2}}{m_{N}^{2} q_{\rm cm}^{\prime}}
    \frac{d \sigma _{\gamma p \to \eta p}}{d \Omega}} .
  \label{eq:TgpTgn}
\end{align}
Empirically, the coefficient $| A_{\rm IS} | / | A_{\rm IS} + A_{\rm
  IV} |$ is estimated as $0.22$--$0.25$~\cite{Weiss:2001yy} with
$\Eglab = 580$--$820 \mev$ from a comparison with theoretical
calculations~\cite{Fix:1997ef, Kamalov:1996qf, Ritz:2000ag}.  In this
study we employ $| A_{\rm IS} | / | A_{\rm IS} + A_{\rm IV} | = 0.22$.

We note that we neglect the phase for this amplitude so that the
amplitude is real. This phase is important when we discuss the
interference between the contributions from the background and the
signal.  We will come back to this point when we discuss the numerical
results in Sec.~\ref{sec:3B}.  For the moment we only mention that
this treatment is satisfactory to estimate how much the $\eta$ meson
is created in the single-step amplitudes, $p 1$ and $n 1$, as the
background.

Next, for the scattering amplitudes of the $\gamma p \to \eprime p$
and $\gamma n \to \eprime n$ reactions, we focus only on their
$s$-wave component because we consider the physics near the $\eprime
N$ threshold.  For the $\gamma p \to \eprime p$ reaction, we have
several data of the cross section~\cite{Williams:2009yj,
  Sumihama:2009gf, Crede:2009zzb} and theoretical
calculations~\cite{Chiang:2002vq, Huang:2012xj, Sakai:2016boo,
  Anisovich:2017pox}.  Here we take the same approach taken in
Ref.~\cite{Sekihara:2016jgx} to determine the $\gamma p \to \eprime p$
amplitude.  Namely, we calculate the scattering amplitude $T_{\gamma p
  \to \eprime p}$ as a function of $\Eglab$ with the formula
\begin{equation}
  T_{\gamma p \to \eprime p} ( \Eglab ) = V_{\gamma 1} 
  + \sum _{j = 1}^{2} V_{\gamma j} G_{j} ( w ) T_{j \, \eprime p} ( w ) ,
  \label{eq:Tg-p}
\end{equation}
with the channel index $i$ [$= 1$ ($2$) for $\eprime p$ ($\eta p$)]
and the center-of-mass energy $w$ fixed as a function of $\Eglab$ as
in Eq.~\eqref{eq:w-Eg}.  The constants $V_{\gamma 1}$ and $V_{\gamma
  2}$ are model parameters and are fixed as
\begin{equation}
  V_{\gamma 1} = 0.348 \gev ^{-1} ,
  \quad 
  V_{\gamma 2} = 0.354 \gev ^{-1} ,
\end{equation}
according to Ref.~\cite{Sekihara:2016jgx}.  These values reproduce the
experimental cross sections with forward proton emission above the
$\eprime p$ threshold~\cite{Williams:2009yj, Sumihama:2009gf}.  As for
the $\gamma n \to \eprime n$ cross section, on the other hand, there
are only few data~\cite{Jaegle:2010jg}.  Nevertheless, as seen in
Ref.~\cite{Jaegle:2010jg}, the value of the $\gamma n \to \eprime n$
cross section near the threshold is similar to that of $\gamma p
\to \eprime p$.  Therefore, we assume that the $\gamma n \to \eprime
n$ amplitude is the same as the $\gamma p \to \eprime p$ one:
\begin{equation}
  T_{\gamma n \to \eprime n} ( \Eglab ) = 
  T_{\gamma p \to \eprime p} ( \Eglab ) .
\end{equation}

\subsubsection{$\gamma d \to \eta d$ scattering amplitude}

Now our task is to fix the scattering amplitude of the $\gamma d \to
\eta d$ reaction, which can be constructed from the amplitudes for
$\eboth$ photoproduction, multiple $\eprime$ scatterings, and
transition to $\eta d$, according to the diagrams in
Fig.~\ref{fig:diag_coh}.

The amplitudes of the single-step scattering, $\mathcal{T}_{p 1}$ and
$\mathcal{T}_{n 1}$, consist of the $\gamma N \to \eta N$ amplitude,
deuteron wave functions in the initial and final states, and the loop
by the nucleon lines.  Therefore, calculating the relative momenta for
the nucleons and integrating them, we can evaluate the amplitude
$\mathcal{T}_{p 1}$ as
\begin{align}
  \mathcal{T}_{p 1} & = T_{\gamma p \to \eta p} ( \Eglab , \, 
  \cos \Theta _{\eta} ) \int \frac{d^{3} q}{( 2 \pi )^{3}}
  \deutwf ( q ) \deutwf ( | \bm{q} - \bm{p}_{d}^{\rm lab} / 2 | )
  \notag \\
  & = T_{\gamma p \to \eta p} ( \Eglab , \, 
  \cos \Theta _{\eta} ) F_{N N} ( p_{d}^{\rm lab} / 2 ) ,
  \label{eq:Tp1}
\end{align}
with the final-state deuteron momentum in the laboratory frame
$\bm{p}_{d}^{\rm lab}$.  The integral part was replaced with the
deuteron form factor $F_{N N}$ in Eq.~\eqref{eq:F_NN}.  We note that
the $\gamma p \to \eta p$ scattering amplitude can be placed out of
the integral by fixing its arguments with external momenta.  Namely,
we can use the same $\Eglab$ as in the free proton target case.  The
$\eta$ scattering angle $\Theta _{\eta}$ can be evaluated from the
Mandelstam variable $t = ( p_{\gamma}^{\mu} - p_{\eta}^{\mu} )^{2}$,
where $p_{\gamma}^{\mu}$ and $p_{\eta}^{\mu}$ are the four-momenta of
the initial photon and the final $\eta$, respectively, as
\begin{equation} 
  \cos \Theta _{\eta} = \frac{\left ( p_{\gamma}^{\mu} - p_{\eta}^{\mu}
    \right )^{2} - 
    m_{\eta}^{2} + 2 q_{\rm cm} \sqrt{(q_{\rm cm}^{\prime})^{2} + m_{\eta}^{2}}}
  {2 q_{\rm cm} q_{\rm cm}^{\prime}} .
  \label{eq:cos_theta_gd}
\end{equation}
The momenta $q_{\rm cm}$ and $q_{\rm cm}^{\prime}$ should be
calculated with Eqs.~\eqref{eq:qcm} and \eqref{eq:qcmp}, respectively.
In some conditions the right-hand side may become more than $1$ or
less than $-1$ because the bound proton is not on its mass shell but
is off-shell due to the Fermi motion.  In such a case we take $\cos
\Theta _{\eta} = 1$ or $-1$, respectively.

In the same manner, we can evaluate the $\mathcal{T}_{n 1}$ amplitude,
and as a consequence we have
\begin{align}
  \mathcal{T}_{p 1} + \mathcal{T}_{n 1} 
  = & \left [ T_{\gamma p \to \eta p} ( \Eglab , \, 
    \cos \Theta _{\eta} ) \right .
  \notag \\ & 
  \quad \left . + T_{\gamma n \to \eta n} ( \Eglab , \, 
  \cos \Theta _{\eta} ) \right ]
  F_{N N} ( p_{d}^{\rm lab} / 2 ) ,
\end{align} 
where the sum of the amplitudes $T_{\gamma p \to \eta p} + T_{\gamma n
  \to \eta n}$ can be evaluated by Eq.~\eqref{eq:TgpTgn}.

\begin{figure}[!t]
  \centering
  \PsfigII{0.19}{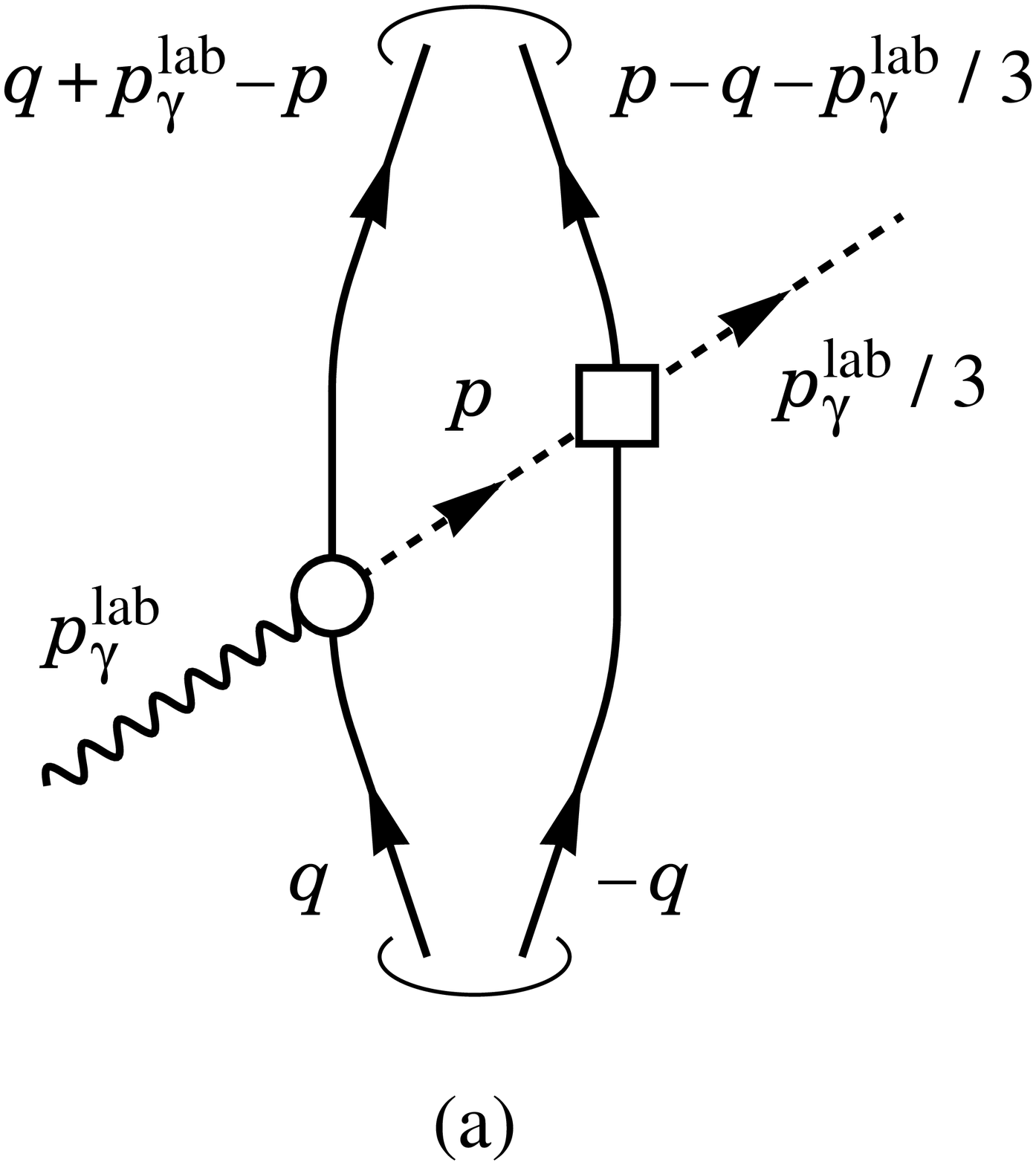} ~
  \PsfigII{0.19}{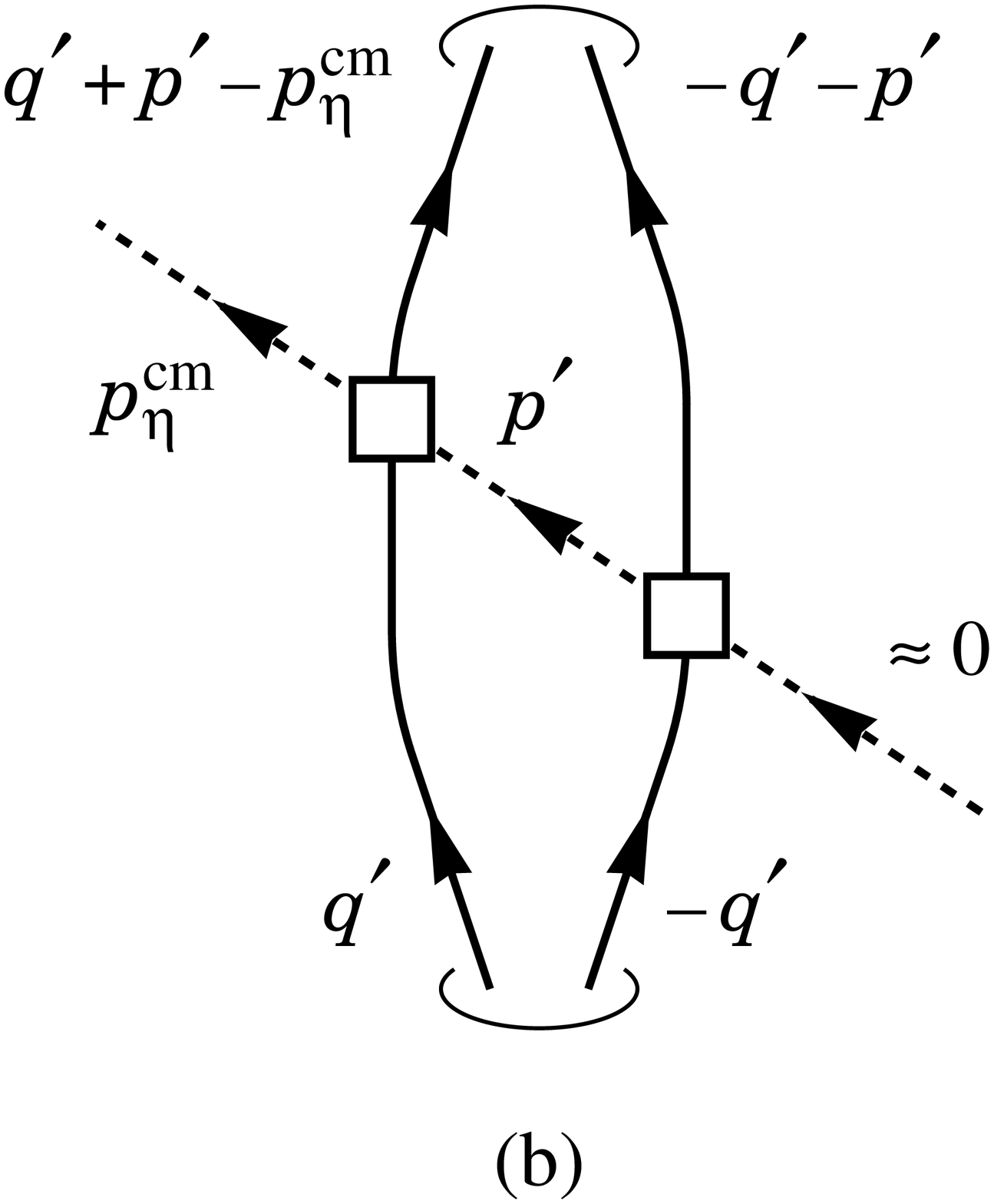}
  \caption{Feynman diagrams for the Green function of the $\eprime$
    propagation (a) after the first collision and (b) before the last
    collision.  The solid, dashed, and wavy lines represent the
    nucleons, $\eboth$ meson, and photon, respectively.  The open
    circles and boxes are not included in the evaluation of the Green
    function.  Three-momenta carried by the particles are shown (a) in
    the laboratory frame and (b) in the total center-of-mass frame.}
  \label{fig:ex_g}
\end{figure}

Next, we fix the double scattering amplitude $\mathcal{T}_{p 2}$.  As
in Fig.~\ref{fig:diag_coh}~($p2$), we construct this with the deuteron
wave functions at appropriate places, $\gamma p \to \eprime p$
amplitude for the first collision, $p n \eprime \to p n \eprime$
amplitude, $\eprime p \to \eta p$ amplitude, and two Green functions
of the $\eprime$ propagation: after the first collision and before the
last collision.

Among them, the two Green functions can be evaluated by using the diagrams
in Fig.~\ref{fig:ex_g}.  For the Green function after the first
collision [Fig.~\ref{fig:ex_g}(a)], the photon momentum should be
shared by $\eprime$ and two nucleons.  Assigning the momenta
$p_{\gamma}^{\rm lab} / 3$ and $2 p_{\gamma}^{\rm lab} / 3$, where
$p_{\gamma}^{\rm lab} = \Eglab$, for the $\eprime$ and deuteron in the
multiple $\eprime$ scattering in the laboratory frame, respectively,
we can evaluate this Green function for the $\eprime$ propagation as
\begin{align}
  G_{\eprime}^{\rm first} & = \int \frac{d^{3} q}{( 2 \pi )^{3}}
  \int \frac{d^{3} p}{( 2 \pi )^{3}}
  \frac{\deutwf ( q ) \deutwf ( | \bm{q} - \bm{p} 
    + 2 \bm{p}_{\gamma}^{\rm lab} / 3 | ) }
       {p_{\eprime}^{0} ( W )^{2} - \bm{p}^{2} - m_{\eprime}^{2} + i 0}
  \notag \\
  & = \int \frac{d^{3} p}{( 2 \pi )^{3}}
  \frac{F_{N N} ( | \bm{p} - 2 \bm{p}_{\gamma}^{\rm lab} / 3 | )}
  {p_{\eprime}^{0} ( W )^{2} - \bm{p}^{2} - m_{\eprime}^{2} + i 0} .
\end{align}
The energy of the mediated meson $p_{\eprime}^{0} ( W )$ was defined
in Eq.~\eqref{eq:peta0}.  For the Green function before the last
collision [Fig.~\ref{fig:ex_g}(b)], we need to bind two nucleons, one
of which has a high momentum $\approx p_{\eta}^{\rm cm}$ coming from
the mass difference between $\eprime$ and $\eta$, to make the
final-state deuteron.  Therefore, the Green function before the last
collision is
\begin{align}
  G_{\eprime}^{\rm last} = & \int \frac{d^{3} q^{\prime}}{( 2 \pi )^{3}}
  \int \frac{d^{3} p^{\prime}}{( 2 \pi )^{3}}
  \frac{\deutwf ( q^{\prime} ) \deutwf ( | \bm{q}^{\prime} + \bm{p}^{\prime}
    - \bm{p}_{\eta}^{\rm cm} / 2 | )}
  {p_{\eprime}^{0} ( W )^{2} - \bm{p}^{\prime \, 2} 
    - m_{\eprime}^{2} + i 0}  
  \notag \\
  = & \int \frac{d^{3} p^{\prime}}{( 2 \pi )^{3}}
  \frac{F ( | \bm{p}^{\prime} - \bm{p}_{\eta}^{\rm cm} / 2 | )}
       {p_{\eprime}^{0} ( W )^{2} - \bm{p}^{\prime \, 2} 
         - m_{\eprime}^{2} + i 0} .
\end{align}
We note that, owing to the integrals, both the Green functions
$G_{\eprime}^{\rm first}$ and $G_{\eprime}^{\rm last}$ do not depend
on the directions of $\bm{p}_{\gamma}^{\rm lab}$ and
$\bm{p}_{\eta}^{\rm cm}$, respectively, and they are functions only of
the center-of-mass energy $W$.

Now we can formulate the scattering amplitude $\mathcal{T}_{p 2}$ as
\begin{align}
  \mathcal{T}_{p 2} = & T_{\gamma p \to \eprime p} ( \Eglab )
  G_{\eprime}^{\rm first} T_{2 2}^{\rm FCA} ( W ) G_{\eprime}^{\rm last}
  T_{2 1} ( w^{\rm FCA} ( W ) ) ,
\end{align}
where $T_{2 1}$ is the $\eprime N \to \eta N$ scattering amplitude in
Sec.~\ref{sec:2A} with its argument $w^{\rm FCA}$ in
Eq.~\eqref{eq:wFCA}.

In a similar manner, we can evaluate the other amplitudes for the
$\gamma d \to \eta d$ reaction:
\begin{align}
  \mathcal{T}_{p 3} = & T_{\gamma p \to \eprime p} ( \Eglab )
  G_{\eprime}^{\rm first} T_{2 1}^{\rm FCA} ( W ) G_{\eprime}^{\rm last}
  T_{2 1} ( w^{\rm FCA} ( W ) ) ,
\end{align}
\begin{align}
  \mathcal{T}_{n 2} = & T_{\gamma n \to \eprime n} ( \Eglab )
  G_{\eprime}^{\rm first} T_{1 2}^{\rm FCA} ( W ) G_{\eprime}^{\rm last}
  T_{2 1} ( w^{\rm FCA} ( W ) ) ,
\end{align}
and
\begin{align}
  \mathcal{T}_{n 3} = & T_{\gamma n \to \eprime n} ( \Eglab )
  G_{\eprime}^{\rm first} T_{1 1}^{\rm FCA} ( W ) G_{\eprime}^{\rm last}
  T_{2 1} ( w^{\rm FCA} ( W ) ) .
\end{align}
Here we note that, because the scatterings of $p 2$, $p 3$, $n 2$, and
$n 3$ take place in $s$ wave, the scattering amplitudes
$\mathcal{T}_{p 2, p 3, n 2, n 3}$ do not depend on the scattering
angle but only on $\Eglab$.  In the full amplitudes, the multiple
scattering amplitude appears as the sum of the $\eprime p n$ and $p n
\eprime$ contributions, i.e., $\mathcal{T}_{p2} + \mathcal{T}_{p3}
\propto T_{2 1}^{\rm FCA} + T_{2 2}^{\rm FCA}$ and $\mathcal{T}_{n2} +
\mathcal{T}_{n3} \propto T_{1 1}^{\rm FCA} + T_{1 2}^{\rm FCA}$.

\subsection{Numerical Results}
\label{sec:3B}

With the scattering amplitudes constructed in the previous subsection,
we can calculate the cross section of the $\gamma d \to \eta d$
reaction.  In the present study the spin components for the photon and
baryons are irrelevant, so we can write the differential cross section
omitting the average and summation of the polarizations as
\begin{equation}
  \frac{d \sigma _{\gamma d \to \eta d}}{d \Omega} = 
  \frac{m_{d}^{2} \, p_{\rm cm}^{\prime}}{16 \pi ^{2} p_{\rm cm} W^{2}}
  \left | \mathcal{T}_{\gamma d \to \eta d} \right |^{2} ,
\end{equation}
where $p_{\rm cm}$ and $p_{\rm cm}^{\prime} = p_{\eta}^{\rm cm}$
denote the momenta of the photon and $\eta$ in the center-of-mass
frame, respectively, and $m_{d}$ is the deuteron mass.

\begin{figure}[!t]
  \centering
  \Psfig{8.6cm}{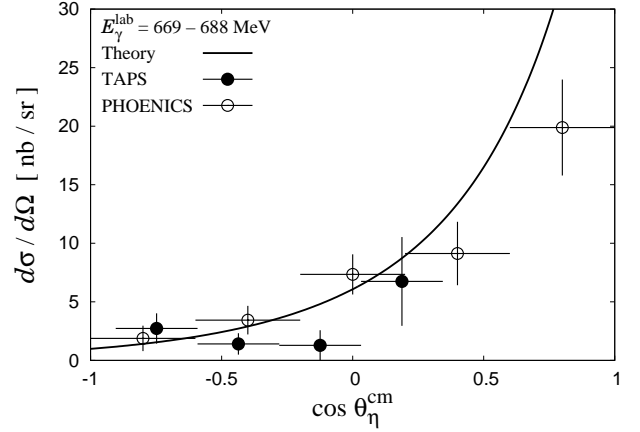}
  \caption{Differential cross section $d \sigma / d \Omega$ for the
    $\gamma d \to \eta d$ reaction with the photon energy $\Eglab =
    669$--$688 \mev$ as a function of $\eta$ emission angle in the
    center-of-mass frame.  The theoretical result is obtained at
    $\Eglab = 680 \mev$.  The experimental data are taken from
    Ref.~\cite{Weiss:2001yy} (TAPS) and from
    Ref.~\cite{HoffmannRothe:1997sv} (PHOENICS).  }
  \label{fig:dsdO_low}
\end{figure}

\begin{figure}[p]
  \centering
  \Psfig{8.6cm}{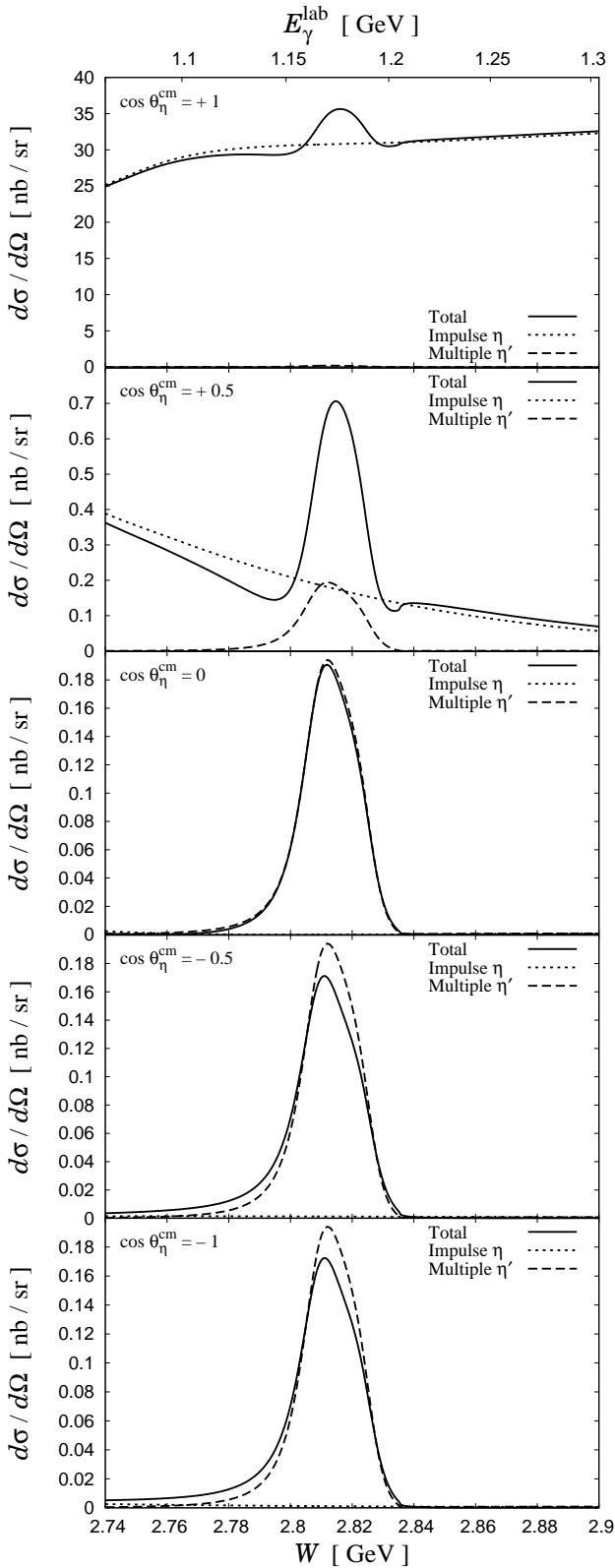}
  \caption{Differential cross section $d \sigma _{\gamma d \to \eta d}
    / d \Omega$ for the $\gamma d \to \eta d$ reaction with scattering
    angles $\cos \theta _{\eta}^{\rm cm} = -1$, $-0.5$, $0$, $+0.5$,
    and $+1$.  The solid lines denote the values for the full
    calculation, while the dotted and dashed lines are contributions
    from the impulse $\eta$ production and the multiple $\eprime$
    scattering, respectively.}
  \label{fig:dsdO_multi}
\end{figure}

Before showing the numerical results in the energy region of the
$\eprime d$ bound-state signal, we demonstrate that the coefficient
for the coherent process $| A_{\rm IS} | / | A_{\rm IS} + A_{\rm IV} |
= 0.22$ [see Eq.~\eqref{eq:TgpTgn}] can reproduce the $\gamma d \to
\eta d$ cross section at slightly above the $\eta$ production
threshold, e.g., $\Eglab = 680 \mev$.  For this calculation, the
$\gamma p \to \eta p$ amplitude is assumed to be constant independent
of both the photon energy and scattering angle and is fitted to
reproduce the cross section summarized by Bonn--Gatchina in the
close-to-threshold region of $\eta$ production off the free proton,
$\Eglab \approx 708 \mev$.  Other terms in the calculation of the
$\gamma d \to \eta d$ amplitude are unchanged.

The numerical result is shown in Fig.~\ref{fig:dsdO_low} with the
photon energy $\Eglab = 680 \mev$.  As one can see from the comparison
with the experimental data at $\Eglab = 669$--$688 \mev$, the cross
section as well as the angular dependence is quantitatively
reproduced.  This means that the present formulation is appropriate
with the coefficient $| A_{\rm IS} | / | A_{\rm IS} + A_{\rm IV} | =
0.22$ and we do not need further normalization factors.  In the
following we use the same value even in the energy region of the
$\eprime d$ bound state.

Now we show the numerical results of the differential cross section
for the $\gamma d \to \eta d$ reaction with the photon energies which
may generate an $\eprime d$ bound state in Fig.~\ref{fig:dsdO_multi}.
The scattering angle is chosen to be $\cos \theta _{\eta}^{\rm cm} =
-1$, $-0.5$, $0$, $+0.5$, and $+1$.  We also plot contributions from
the impulse $\eta$ production ($\mathcal{T}_{p 1} + \mathcal{T}_{n
  1}$) and the multiple $\eprime$ scattering ($\mathcal{T}_{p 2} +
\mathcal{T}_{p 3} + \mathcal{T}_{n 2} + \mathcal{T}_{n 3}$).

Let us consider backward $\eta$ production with $\cos \theta
_{\eta}^{\rm cm} = - 1$.  As one can see from the lowest panel of
Fig.~\ref{fig:dsdO_multi}, the differential cross section is dominated
by the multiple $\eprime$ scattering contribution and the $\eprime d$
bound-state signal is clear as a bump structure with its strength
$\sim 0.2 \nanobsr$.  In backward $\eta$ production, single-step
$\eta$ emission off a bound nucleon is highly suppressed because of a
momentum mismatching between two nucleons in forming a deuteron.  In
this sense, backward $\eta$ production is of interest in searching for
the signal of the $\eprime d$ bound state.  A similar tendency holds
in the scattering angle $\cos \theta _{\eta}^{\rm cm} \le 0$, where
the cross section is dominated by the multiple $\eprime$ scattering
shown in dashed lines.

Next, as the $\eta$ is emitted at more forward angles, the single-step
background contribution becomes much more significant.  At $\cos \theta
_{\eta}^{\rm cm} = +0.5$ the single-step contribution is comparable to
the bound-state signal, and at $\cos \theta _{\eta}^{\rm cm} = 1$ the
single-step contribution is dominant.  However, even at $\cos \theta
_{\eta}^{\rm cm} = +0.5$ and $+1$, we can observe a bump structure
coming from the $\eprime d$ bound state.  At $\cos \theta _{\eta}^{\rm
  cm} = +0.5$ the peak strength is approximately $0.5 \nanobsr$, and
at $\cos \theta _{\eta}^{\rm cm} = +1$ it is about $5 \nanobsr$.

Here we should discuss two ambiguities in our amplitude.  First, in
the formulation of the $\gamma p \to \eta p$ and $\gamma n \to \eta n$
amplitudes, we suppressed the spin component as in
Eq.~\eqref{eq:TgpTgn}.  However, we used these amplitudes only to
estimate the background contribution and to compare it with the signal
strength of the $\eprime d$ bound state.  This background contribution
was found to be negligible in backward $\eta$ production.  Therefore,
we will obtain the bound-state peak in backward $\eta$ production even
if we take into account the spin component rigorously.

Second, as mentioned below Eq.~\eqref{eq:TgpTgn}, we fixed the $\gamma
p \to \eta p$ amplitude as real quantities and did not introduce any
explicit relative phase between the single-step amplitude and multiple
$\eprime$ amplitude.  An important point is that the relative phase
affects the structure for the bound state in forward $\eta$
production.  The bump structure at $\cos \theta _{\eta}^{\rm cm} =
+0.5$ and $+1$ in Fig.~\ref{fig:dsdO_multi} is determined by the
constructive interference between the bound-state formation and the
single-step background contribution.  Such a pattern of the
interference may change owing to the phases of the underlying
reactions.  For instance, if we introduce a relative phase $e^{i \pi}
= -1$, a bump in forward $\eta$ production seen in
Fig.~\ref{fig:dsdO_multi} would become a dip structure due to the
destructive interference.  Besides, the bound-state signal in backward
$\eta$ production will be almost independent of the relative phase
between the single-step amplitude and multiple $\eprime$ one, because
the multiple $\eprime$ scattering dominates the cross section and the
interference is negligible.  In this sense, we may experimentally
discuss the relative phase as well as the strength of the bound-state
signal by investigating the angular dependence of the cross section.

\begin{figure}[!t]
  \centering
  \Psfig{8.6cm}{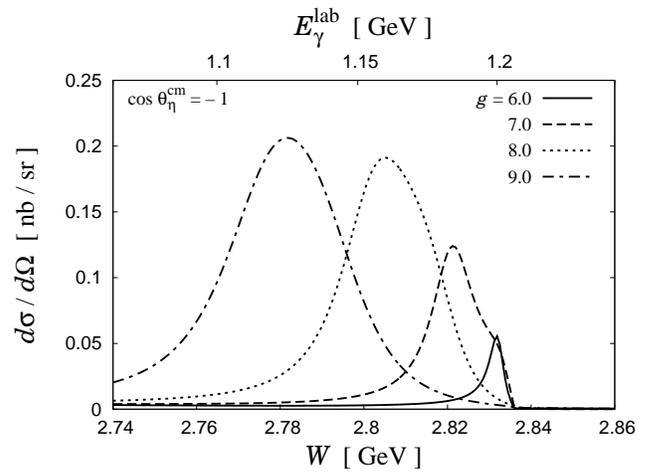}
  \caption{Differential cross section $d \sigma / d \Omega$ for the
    $\gamma d \to \eta d$ reaction with several values of the
    parameter $g$ in the $\eprime N \to \eprime N$
    interaction~\eqref{eq:Vjk}.  The scattering angle is fixed as
    $\cos \theta _{\eta}^{\rm cm} = -1$.}
  \label{fig:dsdO_para_g}
\end{figure}

Before closing this section, we briefly discuss how the signal of the
$\eprime d$ bound state in the $\gamma d \to \eta d$ reaction changes
in case of slightly smaller or larger binding energies of the $\eprime
d$ system.  For this purpose, we vary the strength of the $\eprime N$
interaction via the parameter $g$ in Eq~\eqref{eq:Vjk}, which is the
coupling constant for the $\sigma N N$ vertex.  We plot in
Fig.~\ref{fig:dsdO_para_g} the differential cross section at the
scattering angle $\cos \theta _{\eta}^{\rm cm} = - 1$ with parameters
$g = 6.0$, $7.0$, $8.0$, and $9.0$, which generate the $\eprime d$
bound state with its poles at $2832 - 2 i$, $2821 - 6 i$, $2801 - 11
i$, and $2775 - 17 i \mev$, respectively.  As one can see, when the
coupling constant $g$ is smaller, i.e., the $\eprime N$ interaction is
weaker, the strength of the $\eprime d$ bound-state signal decreases
as well.  On the other hand, a larger coupling constant $g$ brings a
similar strength of the bound-state signal $\sim 0.2 \nanobsr$
compared to that in the case of the original parameter.

\section{Summary}
\label{sec:4}

We theoretically investigated a possibility of binding an $\eprime d$
system by an attractive strong interaction between $\eprime$ and
nucleons.  Thanks to the attractive nature of the $\eprime N$
interaction from the linear $\sigma$ model, which is an effective
model respecting chiral symmetry of QCD, the $\eprime d$ system can be
bound in this model.  With the fixed center approximation to the
Faddeev equation, its binding energy measured from the $\eprime d$
threshold and decay width are $25 \mev$ and $19 \mev$, respectively.

We then proposed the $s$-channel formation of the $\eprime d$ bound
state in the $\gamma d \to \eta d$ reaction at the center-of-mass
energy $\approx 2.8 \gev$, corresponding to the photon energy $\Eglab
\approx 1.2 \gev$.  A clear peak structure with the strength of $\sim
0.2 \nanobsr$ for the signal of the $\eprime d$ bound state was
observed in backward $\eta$ emission, thanks to large suppression of a
background coming from single-step $\eta$ emission off a bound
nucleon.  In addition, the bound-state signal may manifest itself
even in forward $\eta$ emission as a bump or a dip, which depends on
the interference between the bound-state formation and the single-step
background.

This result motivates a new experimental
program~\cite{Fujioka:2017LOI} using the tagged photon
beam~\cite{Ishikawa:2010zza} and the FOREST
detector~\cite{Ishikawa:2016kin} at the Research Center for Electron
Photon Science (ELPH), Tohoku University, Japan.

\begin{acknowledgments}
  This work was partly supported by the Grants-in-Aid for Scientific
  Research from MEXT and JSPS (Nos.~26400287, 
  15K17649
  ).
\end{acknowledgments}

\end{document}